%% file: SIFT-arxiv.tex
\def\year{2021}\relax
\documentclass[letterpaper]{article} 
\usepackage{aaai21}  
\usepackage{times}  
\usepackage{helvet} 
\usepackage{courier}  
\usepackage[hyphens]{url}  
\usepackage{graphicx} 
\usepackage{natbib}  
\usepackage{caption} 
\nocopyright
\usepackage{listings}
\usepackage{algorithm}
\usepackage{enumitem}
\usepackage{amsmath}
\usepackage{booktabs}
\usepackage[noend]{algpseudocode}
\usepackage[para]{footmisc} 
\usepackage{bbm}
\makeatletter
\def\BState{\State\hskip-\ALG@thistlm}
\makeatother
\lstdefinelanguage{json}{
    basicstyle=\normalfont\ttfamily,
    numbers=none,
    numberstyle=\scriptsize,
    stepnumber=1,
    numbersep=8pt,
    showstringspaces=false,
    breaklines=true,
    frame=single,
}
\lstdefinelanguage{py}{
    basicstyle=\normalfont\ttfamily,
    numbers=none,
    numberstyle=\scriptsize,
    stepnumber=1,
    numbersep=8pt,
    showstringspaces=false,
    breaklines=true,
    frame=none
}

\title{Towards Integrating Fairness Transparently in Industrial Applications}
\author{AUTHOR INFORMATION REDACTED}
\author{
    Emily Dodwell,
    Cheryl Flynn,
    Balachander Krishnamurthy,
    Subhabrata Majumdar,
    Ritwik Mitra\\
}
\affiliations{
    AT\&T Labs - Research\\
    \{cheryl,bala,subho\}@research.att.com
}
\begin{document}

\maketitle

\begin{abstract}
Numerous Machine Learning (ML) bias-related failures in recent years have led to scrutiny of how companies incorporate aspects of transparency and accountability in their ML lifecycles. Companies have a responsibility to monitor ML processes for bias and mitigate any bias detected, ensure business product integrity, preserve customer loyalty, and protect brand image. Challenges specific to industry ML projects can be broadly categorized into principled documentation, human oversight, and need for mechanisms that enable information reuse and improve cost efficiency. We highlight specific roadblocks and propose conceptual solutions on a per-category basis for ML practitioners and organizational subject matter experts. Our systematic approach tackles these challenges by integrating mechanized and human-in-the-loop components in bias detection, mitigation, and documentation of projects at various stages of the ML lifecycle. To motivate the implementation of our system---SIFT (System to Integrate Fairness Transparently)---we present its structural primitives with an example real-world use case on how it can be used to identify potential biases and determine appropriate mitigation strategies in a participatory manner.
\end{abstract}

\input{1-introduction}

\input{3-industry}

\input{4-framework}
\input{5-example}
\input{2-background}
\input{6-conc}
\bibstyle{aaai21}
\bibliography{SIFT-arxiv}

\setcounter{section}{0}
\renewcommand\thesection{\Alph{section}} 
\input{appendix}

\end{document}

%% file: 1-introduction.tex
\section{Introduction}
\label{sec:intro}
With the increasingly widespread use of Machine Learning (ML) in our daily lives, numerous ML bias-related problems have surfaced in applications like image
identification \cite{google}, hiring~\cite{amazon}, and targeted advertising~\cite{datta2014automated}. The terms `bias' and `fairness' carry a multitude of implications that may be application-specific, in algorithmic decision making. We use bias to refer to specific quantitative disparities across different demographic groups, addressing ones that can get us closer to an idealized understanding of fairness as appropriate for the use case. Under this premise, most of the academic work till now has been analytical, theoretical, and legal~\cite{Barocas2016big}. Technical methods~\cite{MehrabiEtal19} and toolkits~\cite{aif360} to detect and remedy bias have also been proposed. 

While concerns of unintentional bias have received significant attention in the popular press and research literature, the problem of bias in an industrial setting (with use cases far more diverse than the handful of academic literature exemplars) remains poorly covered. Industry has to serve under-represented communities and not prioritize
the targeting or delivering of services in a discriminating way. It has a responsibility to continually monitor ML processes for bias and mitigate any identified bias to ensure business product integrity, preserve customer loyalty, and protect brand image. However, there are several barriers to a systemic solution to such problems~\cite{Holstein2018}.

We envision an operational framework to tackle fairness concerns at different
stages of an industry ML project workflow. Our system SIFT (System to Integrate Fairness Transparently) enables an industrial ML team to define, document, and maintain a project's bias history. SIFT guides a team via mechanized and human components to monitor fairness issues in all parts of the project. Longitudinally, SIFT lowers the cost for dealing with fairness through reuse of techniques and lessons learned from handling past fairness concerns. To translate our vision into reality, we identify key industry challenges, propose solutions and outline directions of future work for the community of ML practitioners and organizational subject matter experts.

\subsection{Industry-specific challenges} 
Goals and certain aspects of typical industry projects pose unique challenges to ML practitioners. Inspired by recent literature on the needs of industry ML practitioners~\cite{Veale18,Holstein2018,codesign}, we enumerate a few challenges below.

{\bf Principled documentation.}
In the security arena, many attempts were made to codify intrusion detection to prevent attacks. For example, the open-source intrusion detection and prevention system Snort\footnote{\url{https://www.snort.org}} uses both signatures and rules while conducting real-time analysis of Internet traffic to match against attacks. Signatures that are stored, updated, and shared include methods to detect an attack, while rules detect specific vulnerabilities. ClamAV\footnote{\url{https://www.clamav.net}} is an anti-virus toolkit that uses complex routines to detect attacks via user-contributed signatures and allows automatic remote database updates to constantly enhance its functionality.

Just as we keep on finding new vectors of security attacks and instances of privacy leakages, there will continue to be new vectors of potential ML bias arising from data and model reuse, or repurposing of ML approaches for alternate use cases.
Given the data gathering costs and time pressure associated with deploying new projects in large enterprises, this may occur with higher frequency than expected. Mirroring the concept of signatures and rules, throughout this paper we show how the codifying and documenting {\it fairness considerations} may facilitate the planning of future projects.



{\bf Human oversight.}
Risk assessment through human oversight is a key aspect of industry ML workflows. Use-case specific factors to guide such oversight include domain knowledge, bias history of past projects in the enterprise, legal guidelines, and cost considerations of brand impact or regulatory penalties. Similar to privacy, regulation may be proposed in the space of ML. Thus, defensive steps to handle regulatory concerns and transparency mechanisms to demonstrate fairness should ideally be proactively built into the ML development lifecycle. This may not ensure identification and mitigation of {\it all} potential sources of bias, but by including human-level checks the likelihood of fairness concerns impacting an enterprise can certainly be {\it reduced}.

{\bf Cost efficiency.}
In enterprises with multiple application areas, many non-customer facing ML Projects, or those with no demographic/geographic proxies may have no potential bias concerns or human oversight requirements. Managers of such projects would want to quickly move ahead. On the other hand, projects likely to cause serious problems to the enterprise if not handled properly need proper vetting. If fairness concerns are too high without a clear path to mitigation, an early exit may also be needed from the default workflow to reconsider the project design or by collecting additional data. To this end, reuse of contextual knowledge and documentation from past projects not only entails significant cost savings, but also amortizes the cost of maintaining such documentation with time as more projects adopt such practices. 

\subsection{Research questions and contributions}
\label{subsec:contrib}
Given the above goals, we envision SIFT aiming to answer the following overarching questions:

\vspace{1em}
\noindent {\bf Q1:} How should we define, document and update fairness-related information in an industry ML project?

\noindent {\bf Q2:} When and how should we integrate human oversight into project workflows?

\noindent {\bf Q3:} How should we efficiently reuse information on past projects to provide adequate but cost-effective fairness monitoring on an ongoing basis?
\vspace{1em}

\noindent To address the above, the contributions of this vision paper are:
\begin{itemize}[leftmargin=*]
\item Elaborate key industry needs with insights on the key research questions, discuss challenges and propose solutions (Section~\ref{sec:industry}).
\item Introduce the SIFT framework of class objects and pipelines, which directly incorporate both mechanized and human-in-the-loop functions, documents each stage of the bias detection and mitigation process, and utilizes information from prior projects where feasible (Section~\ref{sec:sift}).
\item Show exemplar applications on a use case inspired by real enterprise problems (Section~\ref{sec:example}). 
\end{itemize}

In contrast to project-level methods, tools and artifacts that form the bulk of the current ML fairness landscape (Section~\ref{sec:related}), our vision is that of an enterprise-specific {\it ecosystem} of ML projects. Going beyond project-level bias detection and mitigation, this ecosystem actively enjoins data scientists, program managers, and subject matter experts (from areas including compliance, legal, and PR) to collectively tackle industry-specific challenges through fairness documentation, targeted human oversight, and reuse of practical lessons from past projects. Some of these aspects are certainly common with model and data lineage tracking in ML pipelines \cite{datalineage}, or artifacts such as FactSheets \cite{Arnold18} and Model Cards \cite{Mitchell19}. However these do not preclude----but rather facilitate---the implementation of this broader vision we propose through SIFT. Further, lineage or artifact gathering efforts at project-level can also benefit from the enterprise-level curation aspects of SIFT.

%% file: 3-industry.tex
\section{Addressing Industry Needs}
\label{sec:industry}
As discussed in Section~\ref{sec:intro}, numerous well-known companies have had a variety of ML bias failures. Most of the failures are seen in hindsight to have been preventable with specific oversight provided by human subject matter experts. Lack of balanced representation in hiring is an issue that is well known to HR professionals whose knowledge should have been harnessed to avoid the kind of problems we have seen. Likewise, good public relations and communication expertise present in large organizations could have been leveraged to consider the peril of higher false positives in recognizing faces of minorities. Instead, industry has used ML algorithms' ability to process large amount of training data quickly without considering the potential perpetuation of discrimination. Compliance and legal experts would have alerted to the risk of violating existing laws before deployment of automated decision making in credit scoring customers. Instead of approaching problems on a per-use case basis at the time of deployment, it is essential to embed the right human oversight derived from subject matter experts at the right stages of the ML pipeline. Such a mechanism will guide project managers and data scientists to learn about potential biases and address them in a way that has worked well in other similar projects.

\subsection{Human oversight}
\label{subsec:hog}
A primary requirement of human oversight should be that it is made available in a minimally intrusive manner. Data scientists and project managers should be able to move the project along but be alerted when there is a higher risk of bias creeping in at a particular pipeline. In the absence of a formal process there may be a temptation to ignore bias risk. On the other hand, if the process is too cumbersome data scientists may ignore suggested guidelines. The sweet spot would be to raise alerts based on past projects information and introduce the notion of accountability: the decision maker who moves to the next stage in the pipeline has effectively weighed the risks and decided to proceed. If, on the other hand, a risk of bias is recognized the oversight mechanism should provide a clear way forward.

We identify a number of opportunities for putting humans in the loop of a ML project workflow: (1) data preparation, (2) sensitive category identification, (3) Decision-making steps---whether any past project in the company is similar to the project at hand, to drop a proxy feature from the data, to gather more data, or to retrain a model, and (4) bias risk assessment at {\it multiple stages} of the workflow. Unlike typical bias detection and mitigation methods that can be implemented as code, these steps are not mechanizable by design and require human oversight, but still relate to critical and relevant tasks. They use the expertise of subject matter experts (SME) covering specific areas such as Human Resources, Communications (PR), Legal, Privacy, and Compliance. Most large enterprises have SMEs in these areas who have dealt with similar problems in the past, and can better anticipate inadvertent ML bias. For example, a compliance expert can help ensure that guidelines provided around customer privacy are properly addressed in keeping with local, state, federal, or even international regulations (such as GDPR).

\paragraph{Human Oversight Guides.}
The large number of ML projects in medium and large enterprises coupled with limited number of SMEs in any policy area, could raise questions on the efficacy of the human functions and their potential to slow the pace of development and deployment of an ML project. Such interventions, while inconvenient, safeguard against unintended consequences of rapid deployment both on society and to the enterprise's brand. In light of this, we propose that each SME provides guidance to project teams according to their expertise in the form of a {\it Human Oversight Guide} (HOG)---a question and answer document organized by project stages. 
HOGs are written with data scientists and project managers as target audience (as opposed to high level enterprise-wide principles) and assists them to address potential concerns in a timely manner. 

The curation of SME-guided questions beforehand helps scale the simultaneous use of any oversight {\it mechanism} on a large number of ML projects, as it obviates the need to consult SMEs at every stage. The oversight mechanism should automatically be able to highlight relevant parts of the relevant HOG at different stages in a project, allowing the PM to decide if they need to communicate with the SME for additional clarifications. Before deploying such a mechanism, each SME is presented with a relevant and tailored set of seed questions to assist them in converging on their respective guidelines. If project-specific issues pop up repeatedly that are not covered by a particular HOG, with project managers needing further consultation, the relevant SME would be asked to revise it.


\paragraph{Sample seed questions and answers.}
Each large organization has SMEs in fields relevant to bias detection and mitigation, such as Legal, Privacy, Compliance, Public Relations (PR), and Human Resources (HR). The below seed questions may be shared with a SME in each field.

\begin{enumerate}[leftmargin=*]
\setlength\itemsep{0em}
\item What types of data fall under the purview of the SME?
\item What laws/regulations are there for this data and its use? 
\item When do machine learning projects typically raise concern within the SME’s field? Are there external examples related to ML bias a data scientist should be aware of? 
\item What qualities would enable the data scientist to assess whether a project is low/medium/high risk? Are there ways to mitigate related risks? 
\item Are there data elements that we are not allowed to look at or need specific approval to use in bias mitigation? 
\item What metrics are typically used to evaluate fairness? Is there a standard accepted threshold for each metric?
\item What vetting is done for 3rd party data and what liabilities do we have in using the data? 
\item How would use of data (i.e. descriptive vs. predictive) impact if a project is considered low/medium/high risk? 
\item If a project requires additional data, what are the necessary approval steps? 
\item Are proxy features a concern? Are there cases where proxy features are acceptable and/or appropriate from a business perspective? 
\item What vetting is done for internal models?
\item What vetting is done for 3rd party models?
\item Are there outcomes that always carry risk from the SME’s perspective relative to their field? 
\item Who should a data scientist contact for additional information?
\end{enumerate}

During creation of a HOG, questions may require iteration in consultation with the specific SME to further elaborate upon the question and capture information most relevant from that field for project teams. For example, a question on sensitive category identification for HR may be ``{\it What are the protected classes that may inform identification of sensitive categories?}". A related question for Privacy may be ``{\it Are there privacy concerns in identifying protected classes? Do these concerns vary depending on the data subjects?}" Similarly, additional focused questions may be created for field. A question specific to HR may be ``{\it What laws/regulations are in place for hiring or employee related data?}", while one posed to Privacy may be ``{\it Are there privacy concerns around the reuse of data/models/bias history across business units?}"

The guidance provided by a SME in the form of an answer to each question may be similar to the following:

{\bf Human Resources}

{\bf Q:} {\it What types of data fall under the purview of the SME?}

{\bf A:} There are some key issues surrounding the use of people data, including the importance of having a deep understanding of data elements used.  During modeling there might be a correlation between a school and some outcome, but discrimination in education exists. Performance-related discrimination may be good but race-related is bad. Working with HR provides a clear understanding of people data elements.

{\bf Public Relations}

{\bf Q:} {\it Are there external examples related to ML bias a data scientist should be aware of?}

{\bf A:} Media often identifies cases where individuals don’t seem to be treated fairly and/or seem to have the same opportunities. The output is judged more than the input for, e.g. job applicant screening, better services in some neighborhoods, and best offers and targeted ads going to certain demographics.

\subsection{Bias accountability and transparency}
\label{subsec:biashist}

A company needs to be able to quickly respond to any potential bias concerns raised about an ML model.  In order to do this, documentation on steps taken to reduce bias and the critical team members responsible for the development, deployment, and decision to use the model need to be easily accessible.  This will enable the company to either quickly justify the use of the model or to identify and correct any oversight in their bias review process to avoid a similar situation in the future. Such documentation will also create accountability for the team members involved to ensure due diligence when performing the bias review for an ML model.

\paragraph{Bias history.}
To accomplish this, we propose automatically tracking the ``bias history’’ for every machine learning project as it proceeds through the bias review process. This adds transparency on the exact mechanism of incorporating fairness considerations into ML projects in an enterprise, and has numerous advantages. First, it enables information reuse for future projects and lowers cost across the enterprise in handling bias.  Second, at the end of a project, the bias history would contain the precise nature of any bias detected, the specific steps in the bias review process where it was first detected, the algorithm that helped mitigate it and information on how successful the mitigation was at reducing the bias.  Such transparency helps organizations defend the actions they have taken to address bias.

Opportunities for documenting bias-related steps taken throughout the project not only cover bias detection and mitigation algorithms and their results, but also human oversight considerations leading up to such results. For example, any risk judgements by the PM to kick off a new project that is made from an SME's answer to HOG seed question 4 (Section~\ref{subsec:hog}): ``{\it What qualities would enable the data scientist to assess whether a project is low/medium/high risk? Are there ways to mitigate related risks?}" is useful information. Similar judgements that are worth recording concern the choice of a bias detection metric and its threshold (question 6), or the decisions to drop or keep a proxy feature (question 10). Recording such considerations for posterity serves two purposes: it creates a trail of accountability for the team and company that they have performed due diligence, and also informs similar decisions in subsequent similar projects.

To achieve the above, bias history should record information in a sequential and structured manner. As an exemplar, each sequence element can contain the following fields: the stage of ML pipeline, a list of sensitive feature this element deals with, names of bias detection and mitigation algorithms, mitigation success status, and any additional details. Note that all sequence elements do not need to have every field populated. For example, while taking a human oversight decision on proxy features as mentioned above, bias detection/mitigation algorithm name fields can stay empty. In Section~\ref{sec:sift}, we codify the components to be included and tracked in the bias history, and provide examples of bias history objects in Section~\ref{sec:example}.

\subsection{Information reuse and cost considerations}
\label{subsec:cost}
In our experience there are often commonalities across ML projects within large enterprises. For example, ML projects related to hiring are likely to have similar regulatory requirements, accepted bias metrics, and lists of sensitive attributes that should be considered.  For emerging project areas that require more iterations and new technology to address bias concerns, the ability to learn from them will benefit future projects.  Storing project details and bias histories in a project database that can be queried at the the start of a new project will allow new projects to learn from past similar projects and help streamline the bias review process.  The database can be seeded with existing ML projects at the outset, and will grow as projects are undertaken through this framework.  Once sufficient knowledge from previous projects is collected, this information could be used to create automatic mechanisms to suggest bias metrics and mitigation strategies, thus further reducing the cost and burden of the ML bias review process.

\paragraph{Large collections of ML projects.}
Adoption of an overarching mechanism that integrates human oversight and bias history documentation presupposes prevalence and proper curation of ML projects within an enterprise. This can be ensured by the existence (or creation) of a database of ML projects within a company.
The list of ML projects and corresponding resources within large enterprises is proprietary information.
Nevertheless, valuable proxies demonstrate the growing investment in and importance of ML to the everyday operations and business decisions of various companies. Algorithmia's blind survey of 303 employees across companies representing a range of ML maturity found that 28\% of companies increased their ML budgets by more than 25\% in 2019 \cite{algorithmia}.  Internal ML-focused conferences can draw hundreds if not thousands of employees to present their work and learn about projects across the company. For example, UberML, Uber's annual internal ML conference, drew 500 employees from 50 groups, and an internal Amazon AI/ML event drew thousands~\cite{Robinson19}. Large tech companies with dedicated research organizations including Google, Facebook, Uber, and Amazon regularly publish original research and present at leading ML conferences, and the list of papers published by researchers across these companies number in hundreds. 

With such volume of ML projects, most (if not all) applications have direct or indirect impact on people directly using the products these industries offer. Two of the three top ML use cases, generating customer insights and intelligence, and improving customer experience~\cite{algorithmia}, are inherently customer-facing. Because decisions based on ML projects have the potential to impact billions of external customers, ensuring fairness in product offerings, customer service treatment, and customer analytics is key. Therefore, significant savings are associated with a streamlined process that can detect potential bias across new and existing projects. Among Algorithmia's surveyed companies employing ML solutions, 45\% had models in production for at least 1 year~\cite{algorithmia}. So it is reasonable to assume that documentation of data and methodology either already exists or can be curated and made be available for addition to a bias and model history database. 

\paragraph{Operationalizing a projects database.}
Efforts are already underway to consolidate `AI Failures' in using ML systems; Partnership in AI (PAI) recently created the AI Incident Database~\cite[AIID]{aiincident}. AIID currently crowdsources failure reports, some of which relate to fairness. A properly curated and vetted version of such a database would benefit researchers so they can avoid making those same mistakes while deploying their own ML products. This new effort directly addresses an important requirement we stress upon---using information and insight from past instances of ML implementations.
Their open-source project\footnote{\url{https://github.com/PartnershipOnAI/aiid}} can serve as a starting point for companies to curate databases of their own that are tailored to internal needs and priorities.

Given that ML projects routinely reuse data (often with modifications or sampling) and tweaked models, it is essential to examine the provenance and reliability of datasets, as well as any past applications and concerns. Incentivizing project managers to enforce adequate data and model documentation helps maintain the accuracy of components for future use.

\paragraph{Cost efficiency.}
The cost for maintaining a dedicated database of projects needs to be traded off against financial risk metrics that approximate the cost of negative PR and brand impact. Such explicit analysis of risk vs. return will drive better institutional decision-making. During the early stages of implementing responsible oversight and reuse procedures, there will be few `similar' projects---leading to higher per-project cost for bias detection and mitigation. However, their long-term adoption in an enterprise will more likely than not reduce projected {\it overall} cost. As time progresses, continual use of bias histories of past projects and the human oversight guidelines will be steadily more effective, thus significantly diminishing the per-project cost of bias monitoring to the enterprise. Future research in certain organizational aspects may further drive down cost. For example, it is non-trivial to quickly determine how changes to data or model may impact bias at different downstream stages of ML lifecycle. Research remains to be done to compartmentalize the potential impact of such modifications. Effectively, we would like to limit the parts of the lifecycle that would be impacted by any modifications to the input data, changes to the model, application to new use cases etc. For specific use cases, possible ways to mechanize parts of human oversight procedures are also worth exploring.

%% file: 4-framework.tex
\section{The SIFT Framework}
\label{sec:sift}
In this section, we introduce the design primitives of our proposed system SIFT, along with its key components. We assume the existence of a database of existing ML projects in the company with the structure below; we denote this database by Pdb. We refer to the team working on the ML project as the SIFT user. Note that this class structure is declarative in nature---we do not require that every project have all class components available at every stage of a project (for example a new project may not initially know the relevant sensitive features), or the use of specific detection and mitigation methods, tools or artifacts. 

\subsection{SIFT classes}
The SIFT framework has four classes: project, data, bias history and
model history. The {\bf\textsc{sift project}} class is the top-level class
for the system with all project-related information. We assume that a SIFT
user starts a new project with (1) a {\bf name}, (2)
a {\bf description} summarizing the background
and objective, and (3) {\bf data location} of the database 
that stores data for the project (a remote directory, network drive, or URL). Based on these three input arguments, the user initiates a
{\bf\textsc{sift project}}. Other components of a {\bf\textsc{sift project}} are:
{\bf project ID}; {\bf\textsc{sift data}}, {\bf\textsc{bias history}}, and {\bf\textsc{model history}}: objects of different classes described below; {\bf metadata}; {\bf model flow}: `Standard' or `Custom' bias-aware model building strategy (Section~\ref{subsec:pipelines}); {\bf similar projects}: pointers to similar projects in Pdb; {\bf older versions}: pointers to previous versions of the project in Pdb; and {\bf timeout}: set to a company-specified time frame for terminated projects to be removed from the project database; defaults to \textsc{NULL} otherwise.

The {\bf \textsc{sift data}} class stores information about data used in the project. This includes raw data, data definitions, a list of feature variables, the target variable, and predicted outcomes from any existing pre-built model.
This class also contains the list of sensitive features relevant to the project, and a summary variable with any additional information relevant to bias investigation. While we consider three such categories of information (sparse groups, proxy features, and marginalized groups) in this paper, specific project teams may decide to include other categories. See Appendix~A.1 for a full list of {\bf \textsc{sift data}} class components.

The {\bf \textsc{bias history}} class is a novel and major part of SIFT, and is used to track each fairness-related stage in the ML workflow. The components of this primitive are: {\bf step}: counter capturing the place in the sequence of bias and mitigation tasks performed; {\bf pipeline}; {\bf bias features}: sensitive features under consideration in the current step, {\bf bias detection function},
{\bf bias mitigation function}; {\bf mitigation success status}; {\bf details}: additional information, such as bias investigation results or actions taken by the SIFT user. Steps in the bias history are added to document each stage of the bias detection and mitigation process. In Table~\ref{tab:pipe}, we mark the stages in each SIFT pipeline that adds a step to the bias history sequence. For methods and mechanisms to access the current step, add components to the current step, or add a next step in a {\bf \textsc{bias history}}, please refer to Appendix A.2.

As mentioned in Section~\ref{subsec:biashist}, bias history adds transparency on exactly {\it how} fairness is weaved into the ML lifecycle in an enterprise. The success status is key in deciding if a recommended mitigation algorithm should be reused. Note that this value is not dispositive; a different mitigation algorithm might work better for a different use case or project. However it is still useful information that is traditionally not tracked in the ML lifecyle. For reuse purposes, {\bf\textsc{bias history}} objects of similar projects are returned as a part of query result, indicating that the manner of a project-level bias resolution is visible to anyone in the future.


The {\bf \textsc{model history}} class tracks the history of the ML model
through development and training. This class includes information on the training and test sets, the fitted model object, the performance metric(s) used to evaluate the model, and its deployment status at each stage of the modeling process. The class components are described in Appendix A.3. 

\subsection{SIFT pipelines}
\label{subsec:pipelines}
\begin{figure}[h]
\centering
\includegraphics[width=1\columnwidth]{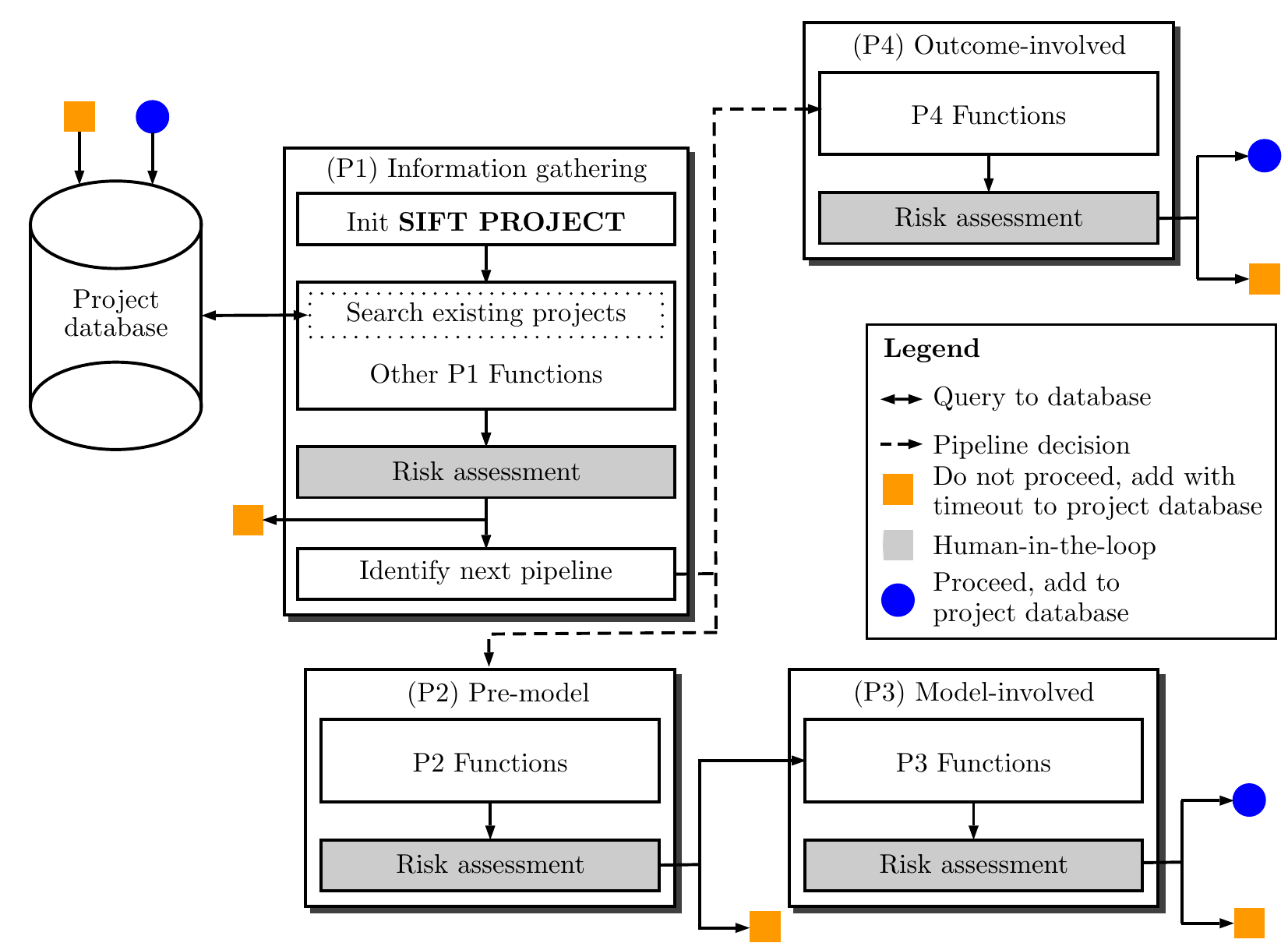}
\caption{The four pipelines of SIFT.}
\label{fig:sift}
\end{figure}

SIFT assists the user through four pipelines (Figure~\ref{fig:sift}). Progress through the pipelines is {\it not} sequential: After initiating a new project in Information gathering, users move to Outcome-involved if the model is already deployed, otherwise to Pre-model followed by Model-involved pipelines. We briefly describe each pipeline below and list their component stages in Table~\ref{tab:pipe}; details on each stage are available in Appendix B.

\begin{table}[h]
\centering
\scalebox{0.9}{\begin{tabular}{l|l}
\toprule
{\bf Pipeline}  & {\bf Stages}    \\\hline
& \parbox[t]{5cm}{Search in Pdb for similar projects} \\ 
Information & Verify similarity$^{H}$  \\
gathering & Identify sensitive categories$^{H,B}$ \\
& Risk assessment$^{H,B}$ \\
& Identify next pipeline \\\hline
& Prepare data$^{H}$ \\
& Detect sparse group$^{B}$ \\
& Decide if more data is needed$^{H}$ \\
Pre-model & Detect proxy features$^{B}$ \\
& Decide whether to drop proxy features$^{H,B}$ \\
& Detect marginalized groups$^{B}$ \\
& Risk assessment$^{H,B}$ \\\hline
& Pre-processing detection$^{B}$ \\
& Pre-processing mitigation$^{B}$ \\
& Train model \\
Model & Post-processing detection$^{B}$ \\
-involved & In-processing mitigation$^{B}$ \\
& Post-processing detection$^{B}$ \\
& Post-processing mitigation$^{B}$ \\
& Risk assessment$^{H,B}$ \\\hline
& Detect covariate shift$^{B}$ \\
Outcome- & Decide if retraining needed$^{H}$ \\
involved & Post-processing detection$^{B}$ \\
& Post-processing mitigation$^{B}$ \\
& Risk assessment$^{H,B}$ \\\bottomrule
\end{tabular}}
\caption{Stages in each SIFT pipeline. $H$ and $B$ in superscripts for a stage signify Human-in-the-loop and bias-history updates, respectively. 
}
\label{tab:pipe}
\end{table}

\noindent{\bf P1. Information gathering:} To begin with, we look for similar projects within the enterprise that may help, for example, to identify
sensitive feature(s). Based on the risk assessment done using the
information collected here, the user may conclude that no fairness
concerns exist and exit SIFT. Referring back to the sample HOG questions in Section~\ref{subsec:hog}, note that questions 2--5 and 7 refer to risk assessment in P1.

\noindent{\bf P2. Pre-model:} If the ML model is not deployed, we move to
the Pre-model pipeline. This provides a framework to explore potential
issues in the data set that could result in a biased model. Considerations include detection of sparse groups, proxy features, and marginalized groups. For P2 stages, question 9 in sample HOG questions pertains to deciding if more data is needed, question 10 concerns dropping or keeping proxy features, and question 8, 10 are useful in risk assessment.

\noindent{\bf P3. Model-involved:} This pipeline guides the user through building an ML model with fairness considerations. If bias is detected, mitigation can be done via one or more pre-processing \cite{Calmon2017,Feldman2015}, in-processing \cite{Celis2018,Zhang2018}, or post-processing algorithms \cite{Hardt2016, Pleiss2017}.  SIFT defaults to a sequential implementation of these strategies if bias is detected; we call this the {\it standard flow} of P3 (See Table~\ref{tab:pipe} for its stages). Instead, a company may also choose to leverage its in-house ML tools, domain-specific methods, or novel mitigation algorithms in a {\it custom flow}.

\noindent{\bf P4. Outcome-involved:} If working with a deployed model, we move directly from P1 to here and check the outcome variable from the deployed model for bias. A post-processing algorithm may be implemented if bias is detected. If the original training data or summary statistics about the data are available, then we also check for distributional changes to the underlying data that could result in bias and require re-training. SIFT further allows for periodic monitoring of a deployed model with comparisons against the original training data through a list of pointers to prior iterations of the project. Note that question 13 in the sample HOG questions helps during risk assessment in P4.


%% file: 5-example.tex
\section{Example use case}
\label{sec:example}
We now show how SIFT can proactively address fairness concerns in two examples covering a representative use case, along with their {\bf\textsc{bias history}} objects. We give all {\bf\textsc{bias history}} fields in the first example (Listing 1), and only nonempty fields in the second for brevity.

\subsection{Motivation}
As targeted advertising has become standard in the digital landscape, concerns of unethical or illegal advertising have also arisen. Historically marginalized groups have lost visibility into information in ads related to high-paying jobs~\cite{datta2014automated, datta2018}. Such discrimination may be unintentional on the part of the advertiser or the ad platform, but nevertheless does occur when targeting systems and ad delivery algorithms are applied without careful evaluation of unexpectedly introduced bias~\cite{lambrecht2018} along the way \cite{celis2019}.

Suppose a company wants to identify customers likely to be early adopters of a new service being rolled out among its existing customer base to receive exclusive promotional discounts. To build an early adopter model, a project team surveys a small sample of customers on their likelihood of service adoption. Based on the responses, the team constructs a binary input feature indicating whether each customer is likely to be an early adopter, and as input features use their marketing data purchased externally---which is available for the entire customer base, and includes demographic information and consumer segmentation data constructed from social media, online browsing, and purchases.

At this stage, the team would like to know the following:
\begin{enumerate}[label=\alph*)]
\setlength\itemsep{0em}
    \item If anyone else in the company worked on similar projects?
    \item If so, what were the sensitive features in their project, and what guidance do these past projects offer on that?
    \item Would legal and compliance allow use of these features?
    \item Does the data contain any features that may act as proxies for any sensitive feature?
    \item Are any groups determined by the sensitive attribute under-represented by the survey respondents?
    \item How to detect and mitigate any bias in the model outcomes with respect to the sensitive feature based on the collected data?  

\end{enumerate}
While most of the current related work (Section~\ref{sec:related}) focus on the last three questions, structured guidance on each of the rest empower the team to move forward to the next task and through the ML workflow---where guidance {\it is} available through existing technical tools---with more confidence.

\subsection{Implementation}
\begin{figure}[h]
\includegraphics[width=\linewidth]{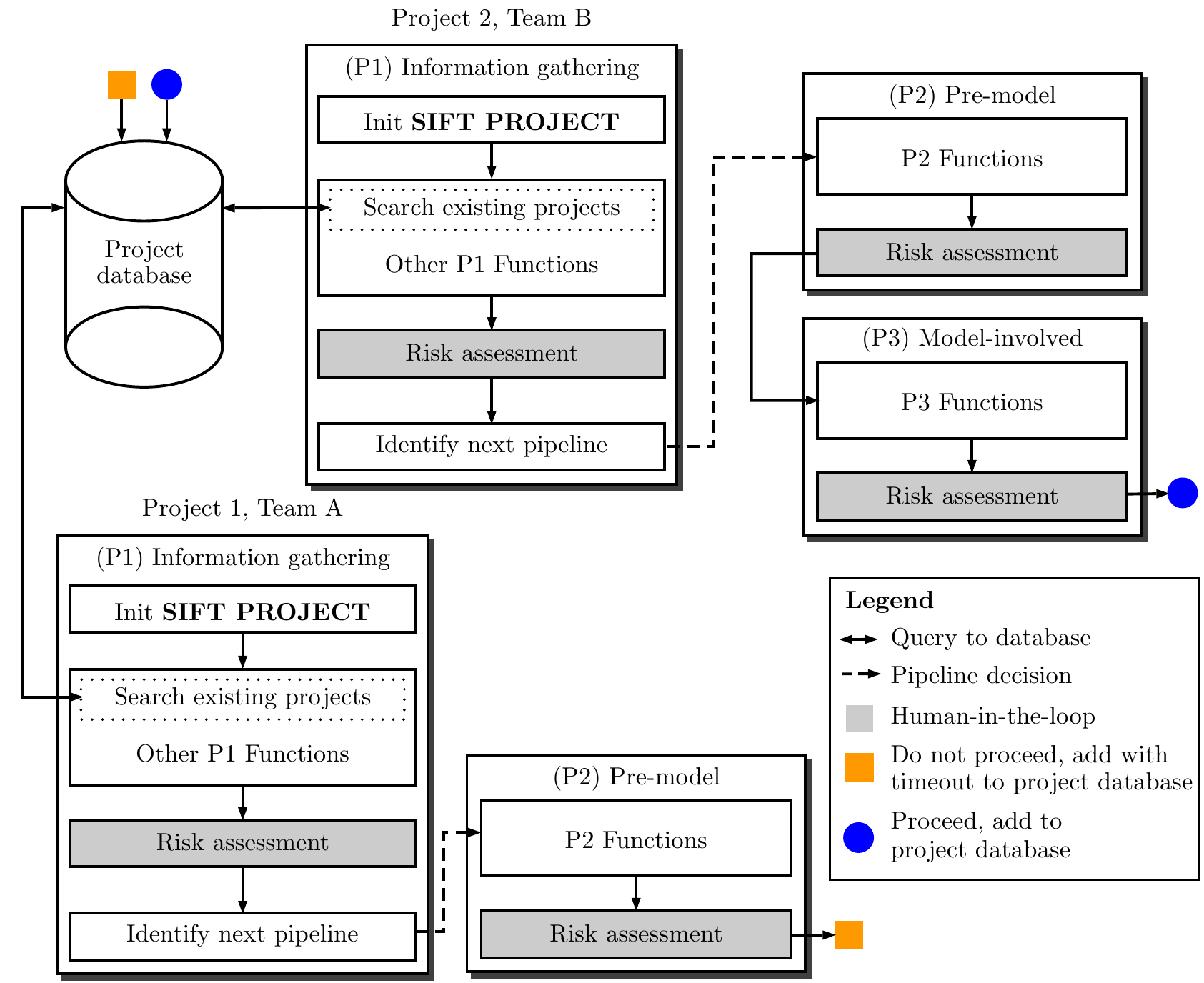}
\caption{SIFT flow for the two marketing-related projects}
\label{fig:ad}
\end{figure}

Below we describe two sample projects under this setup and their steps through the SIFT pipelines. We summarize these steps  in Figure~\ref{fig:ad}. Note that both projects skip the Outcome-involved pipeline because neither involve a deployed model.

As data for these projects, we use the demographic data from the UCI
Adult dataset \cite{uci} 
and simulate 50 binary features designed to represent consumer
segmentation data.
We simulate ${\bf y}$ as a
binomial random variable with probability depending on a subset of the
features; simulation details are provided in Appendix~C.
A small subsample is selected that contains only 5\% non-white samples in Project 1.  The full dataset is used in Project 2.

%

\noindent{\bf Project 1}

{\bf Information gathering.} Team A initializes a
{\bf\textsc{sift project}} object with project id Svc2020. The corresponding {\bf\textsc{sift data}} object is populated with the target variable, demographic data, and consumer segmentation data.  Then a search for existing projects obtain a list of similar projects in Pdb, and a human user verifies the similarity with the current project. The team connects with other business units working on the identified similar projects and learns from their subject matter experts (SME).
After considering the collected information, marital status, race, and sex are determined to be sensitive features.

The company would prefer to avoid the 
bias of offering discounts for the service disproportionally to any of
the demographic subgroups identified by the sensitive features.  
After consulting legal and compliance, human risk assessment determines that the project should proceed through the SIFT system due to a risk of potential bias. Prior projects used in the company's standard model flow, and there are no
additional cost or computational constraints for this project.  Thus,
the standard model flow is selected.  Disparate impact is the bias
detection metric specified in the standard flow for marketing
applications.  The specified fairness range is $(0.8, 1.2)$, outside
of which bias is detected. No model is deployed, thus
the next the next pipeline is identified as Pre-model, and the {\bf\textsc{bias history}} and {\bf\textsc{model history}} objects are initialized.

{\bf Pre-model.} After data preparation,
a sparse group detection algorithm checks that each subgroup defined by each
sensitive feature makes up at least 10\% of the data samples, and identifies an under-representation of non-white customers in the dataset. The SIFT user decides to collect additional data for the project. The project is terminated, with
{\bf\textsc{bias history}} updated accordingly.  The project is added
to the Pdb with a {\bf \textsc{sift project}.timeout} set to 1-year pursuant to the company's settings.

{\scriptsize
\begin{lstlisting}[caption={Bias history of Project 1},label={lst:hist1},language=json]
{"bias_history": [
  { 'step' : 0,
    'sift_pipeline' : 'Information gathering',
    'bias_features' : '',
    'bias_detection_function' : '',
    'bias_mitigation_function : '',
    'mitigation_success_status' : '',
    'details' : 'Risk assessment indicates project should proceed through SIFT.'},
  { 'step' : 1,
    'sift_pipeline' : 'Pre-model',
    'bias_features' : '{'sex','race','marital_status'}',
    'bias_detection_function' : 'computeSampProportion',
    'bias_mitigation_function : '',
    'mitigation_success_status' : '',
    'details' : 'Get additional data.'},  
  { 'step' : 2,
    'sift_pipeline' : 'Exit SIFT',
    'bias_features' : '',
    'bias_detection_function' : '',
    'bias_mitigation_function : '',
    'mitigation_success_status' : '',
    'details' : 'Team will collect additional data.  Project terminated and added to project database.'}]}
\end{lstlisting}
}


\noindent{\bf Project 2}

{\bf Information gathering.} Six-months later, after collecting additional data from a new survey, Team B initializes a {\bf\textsc{sift project}} object with project id NewSvc2020, populating {\bf\textsc{sift data}} with the new data. A search for existing project and human vefication of similarity obtains a list of similar projects in Pdb, along with the old project Svc2020. A pointer to Svc2020 is added to the list in {\bf \textsc{sift project}.older versions}.  Based on the information available from Svc2020, the team quickly determines that NewSvc2020 should proceed through SIFT. Project Svc2020's sensitive features and model flow selection are copied into the new project. No model is deployed, thus the next stage is set as Pre-model, and new {\bf\textsc{bias history}} and {\bf\textsc{model history}} objects are initialized.

{\bf Pre-model.} This time no sparse groups are identified in the new dataset.
For proxy feature detection, a Chi-Square test for independence on each (sensitive feature, non-sensitive feature) pair is performed, which compares the $p$-value against a Bonferroni-corrected threshold of $0.01 / m$, $m$ being the number of non-sensitive features. No proxy features are identified. Lastly,
Disparate Impact is computed between $y$ and each sensitive feature (Table~\ref{t:mkt-bias-results}, Column 2). All results are within the fairness range of (0.8, 1.2), so no marginalized groups are detected. The pipeline updates the {\bf\textsc{bias history}} after each of these steps with the corresponding
bias detection algorithm and result. The SIFT user decides to proceed
to the Model-involved pipeline after another human risk assessment.

\begin{table}[h]
\caption{Bias detection metric results for Project 2}\label{t:mkt-bias-results}
{\footnotesize
\begin{tabular}{|c|c|c|c|}
\hline
& \multicolumn{3}{|c|}{Disparate Impact}  \\
\hline
Sensitive Feature & $y$ &Original Model & Debiased Model \\
\hline
marital\_status & 0.85 & 0.82 & 0.83 \\
race & 0.96 & 0.97 & 1.00 \\
sex & 0.84 & 0.79 & 0.88 \\
\hline
\end{tabular}}
\end{table}

{\scriptsize
\begin{lstlisting}[caption={Bias history of Project 2},label={lst:hist2},language=json]
{"bias_history": [
  { 'step' : 0,
    'sift_pipeline' : 'Information gathering',
    'details' : 'Risk assessment indicates project should proceed through SIFT.'},
  { 'step' : 1,
    'sift_pipeline' : 'Pre-model',
    'bias_features' : '{'sex','race','marital_status'}',
    'bias_detection_function' : 'computeSampProportion',
    'details' : 'No sparse groups detected.'},  
  { 'step' : 2,
    'sift_pipeline' : 'Pre-model',
    'bias_features' : '{'sex','race','marital_status'}',
    'bias_detection_function' : 'computeChiSqTest',
    'details' : 'No proxy features detected.'}, 
  { 'step' : 3,
    'sift_pipeline' : 'Pre-model',
    'bias_features' : '{'sex','race','marital_status'}',
    'bias_detection_function' : 'computeDispImpact',
    'details' : 'No marginalized groups detected.'},
  { 'step' : 4,
    'sift_pipeline' : 'Model-involved',
    'bias_features' : '{'sex','race','marital_status'}',
    'bias_detection_function' : 'computeDispImpact',
    'bias_mitigation_function' : 'adversarialDebiasing',
    'mitigation_success_status' : 'TRUE',
    'details' : 'Bias detected in model outcome. In-processing strategy implemented.'},
  { 'step' : 5,
    'sift_pipeline' : 'Exit SIFT',
    'details' : 'Project scheduled for deployment and added to project database.'}]}
\end{lstlisting}
}

{\bf Model-involved.} The team proceeds to split the data evenly into a training and test set, and trains a logistic regression model. The model has a test-set accuracy of 77.6\%.  The {\bf\textsc{model history}} object is updated accordingly. Disparate Impact is calculated on the predicted outcomes for the test set, as set by the company's standard flow settings (Table~\ref{t:mkt-bias-results}, Column 3).  The results show that bias is detected on the basis of the sensitive feature {\bf `sex'} since 0.79 is outside of the fairness range.

Next, as part of standard flow, an in-processing mitigation applies Adversarial Debiasing \cite{Zhang2018} to correct the detected bias.  The {\bf\textsc{model history}} object is updated accordingly. A final disparate impace checkconfirms the absence of any bias in the predicted outcomes of the debiased model (Table~\ref{t:mkt-bias-results}, Column 4).  The test-set accuracy of the debiased model is 76.2\%.  These steps are recorded in {\bf\textsc{bias history}}.

A final human risk assessment confirms only a minimal drop in accuracy
between the original and debiased models and shares {\bf\textsc{bias history}}
with legal and compliance, who confirm that the bias has been addressed.
The {\bf\textsc{sift project}} object is updated now and returned to the ML
projects database as scheduled for deployment.

Given the importance of time-to-market in campaign deployment, copying the information collected by Team A and other similar projects was an important time saving measure for Team B.


%% file: 2-background.tex
\section{Related Work}
\label{sec:related}



Research in bias and fairness till now can be divided into three broad categories.

{\bf Methodology.}
Depending on the use case and modeling objectives, several sources of
bias and discrimination may exist in the data. A recent
enumeration~\cite{MehrabiEtal19} lists 23 types of bias
and 6 types of discrimination associated with ML models, exposing
three types of fairness concerns: individual, group, and subgroup
fairness. Research on bias detection and mitigation methodology
include methods aimed at the pre-, in- or post-processing parts of
an ML project using classification or regression modeling \cite{MehrabiEtal19}. Fair versions of other techniques such as
clustering~\cite{Backurs19}, community detection~\cite{Mehrabi19},
PCA~\cite{Samadi18}, and causal models~\cite{Zhang17} have also been proposed. 

{\bf Tools.}
AI Fairness 360~\cite[AIF360]{aif360} 
is a well-known tool that packages bias detection and
mitigation methods in the literature for reuse. Among other similar
packages, which are mostly open-source, Aequitas~\cite{Stevens18}, Fairness Measures~\cite{Zehlike17}, FairML~\cite{Adebayo16}, FairTest~\cite{Tramer17}, and Themis~\cite{Galhotra17} offer
bias detection, while Fairlearn~\cite{fairlearn} and Themis-ml~\cite{Bantilan18} offer detection and mitigation through expandable platforms.
LinkedIn Fairness Toolkit~\cite[LiFT]{lift} was recently released as an open-source framework that 
can handle web-scale ML problems. 

%
{\bf Artifacts.}
The above tools and methods focus on the technical requirements for integrating bias metrics and mitigation algorithms into ML projects.  However, there are several other challenges---including the ones discussed earlier---for integrating fairness considerations into industry applications~\cite{Holstein2018,Veale18}. These pertain to lack of guidance on data collection, blind spots, use case diversity and the need for human oversight. Motivated by such needs, recently a number of data and model documentation artifacts have been proposed to enable transparency in industry ML processes, such as FactSheets~\cite{Arnold18}, Datasheets~\cite{Gebru18}, and Model Cards~\cite{Mitchell19}. These can be adapted to detect bias concerns in the data or at different stages of the ML workflow. In human oversight, implementation of internal algorithmic audits \cite{smactr} and co-designed fairness checklists \cite{codesign} help ensure that deployed ML models conform to company values and principles.

To summarize, existing research on fairness has mostly focused on mechanisms to monitor and tackle fairness concerns in a dataset/model/project, but not much on exactly how to harness this knowledge and domain expertise {\it across projects} towards the understanding of the fairness landscape inside a company. We aim to address this gap through SIFT, while allowing project teams to be flexible in choosing bias monitoring tools and artifacts optimal for their own team and enterprise. Data scientists can get started on bias detection and mitigation with methods implemented in the open-source libraries like AIF360 or LiFT, then code up other methods on an on-demand basis. Existing implementations of DataSheets, Factsheets, or Model cards in the enterprise can inform or enrich components of a {\bf\textsc{sift project}} object. Human risk assessment steps can be amply facilitated by structured algorithmic audits \cite{smactr} and fairness checklists \cite{codesign}.

%% file: 6-conc.tex
\section{Conclusion}
\label{sec:sec6}
In this vision paper, we have shown how bias detection and mitigation in ML may be done in an enterprise setting in a holistic, transparent, and accountable manner. Industry's technical challenges include diversity of use cases, datasets, and audiences. Further, companies have to ensure that no demographic bias arises even well after deployment while adhering to policy recommendations and meeting compliance requirements. While SIFT---our proposed framework---does not handle all these problems, it shows how a large class of ML projects can follow a structured approach to reduce chances of bias going undetected until it is too late. 

%% file: appendix.tex
\section*{Appendix}

\section{SIFT Class components and methods}
\label{app:sift_class_details}

We provide additional information about the {\bf\textsc{sift data}} class
components, {\bf\textsc{bias history}} methods, and the {\bf\textsc{model history}}
class components here.

\subsection{SIFT data class components}\label{app:sift_data_details}
The list of components of the {\bf\textsc{sift data}} class are:
\begin{itemize}[leftmargin=*]
\setlength\itemsep{0em}
\item {\bf raw data} -- a connection to the data for the project, such as a dataframe or a connection to a distributed file system that provides access to the data,
\item {\bf data definitions} -- a dictionary with definitions for each variable in the dataset,
\item {\bf y} -- a variable with the name of the response variable,
\item {\bf X} -- a list with the names of the predictors,
\item {\bf outcome} -- a connection to the predicted outcomes from the
  ML model; This will be pre-populated if the model is deployed else
  populated after model training,
\item {\bf sens features} -- a list with the names of the sensitive features that contain categories that might suffer from biased treatment,

\item {\bf sens features summary} -- a dictionary of dictionaries
  where each key denotes a specific characteristic relevant to
  bias-investigation, such as sparse groups, proxy features, or
  marginalized groups, and each corresponding value defines a
  dictionary with keys denoting sensitive features as identified in
  {\bf sens features} and values being lists describing the
  corresponding characteristics.
\end{itemize}
As a concrete example of the {\bf sens features summary}, assume income is a sensitive feature and we are trying to identify its proxy features, which are education and race. This will be represented in the {\bf sens features summary} as
{\scriptsize
\begin{lstlisting}
sens_features_summary = {
  proxy_features: {income: [education, race]}
}
\end{lstlisting}}
\noindent If multiple features were to jointly act as proxy for a sensitive feature they could be represented as a list within the list of proxy features. 

The components of the {\bf\textsc{sift data}} class may be populated manually
or via mechanizable extraction functions that take the {\bf data location}
and project description as arguments.

\subsection{Bias history class methods}
\label{app:bias_hist_methods}
During the course of the project, the {\bf\textsc{bias history}} object is
updated through methods associated with the class to reflect the bias
detection and mitigation steps performed.  To get the current step we
use the method:

{\scriptsize
\begin{lstlisting}[language=py]
def getLatestStep(sift_project):
  return sift_project.bias_history[-1]['step']
\end{lstlisting}
}
\noindent Components are added to the current step to track bias
investigation at each step using the class method:
{\scriptsize
\begin{lstlisting}[language=py]
def insertBiasHistoryAt(sift_project, insert_at, **kwargs):
  if (insert_at > sift_project.getLatestStep()):
    return 'cannot insert outside current history range'
  #	 fill in components and corresponding values from **kwargs
  for key, value in kwargs.item():
    if key in list(sift_project.bias_history[0].keys()):
      sift_project.bias_history[insert_at][key] = value
    else:
      print(f'{key} is not an attribute of bias history')
\end{lstlisting}
}

\noindent Lastly, the next step in the bias history is added using the
class method:
{\scriptsize
\begin{lstlisting}[language=py]
def addBiasHistoryStep(sift_project, **kwargs):
  # append new step  
  sift_project.bias_history.append( 
    {'step'                    :sift_project.getLatestStep() + 1, 
    'sift_pipeline'            :None, 
    'bias_features'            :None,
    'bias_detection_function'  :None, 
    'bias_mitigation_function' :None,
    'mitigation_success_status':None,
    'details'                  :None})
  
  #	 fill in components and corresponding values from **kwargs
  for key, value in kwargs.item():
    if (key != 'step') and 
      (key in  list(sift_project.bias_history[0].keys())):
      insert_at = sift_project.getLatestStep()
      sift_project.bias_history[insert_at][key] = values
\end{lstlisting}
}

\subsection{Model history class components}\label{app:model_hist_details}
The SIFT model history class keeps track of the modeling efforts.  The list of components of the {\bf\textsc{model history}} class are:
\begin{itemize}[leftmargin=*]
\setlength\itemsep{0em}
\item {\bf step} -- a counter capturing the place in the sequence of modeling tasks performed,
\item {\bf seed} -- the random seed used in the modeling process to ensure reproducibility,
\item {\bf train index} -- the set of indices in the raw data used for training the ML model,
\item {\bf test index} -- the set of indices in the raw data used for testing the ML model,
\item {\bf fitted model} -- includes the loss function to be optimized and its value for the fitted model, the tuning parameters, and the estimated model,
\item {\bf perf metric} -- a dictionary that includes the name of the performance metric used to evaluate the model and the performance metric value for the test-set; For example, the performance metric for a classification problems could be accuracy, precision, recall, etc., 
\item {\bf is deployed} -- a flag indicating if the model is deployed.  If \textsc{true} at step 0, then this indicates that the project was initiated with an earlier model already deployed.
\end{itemize}

\section{SIFT pipeline details}
We provide descriptions of stages of the four SIFT pipelines, and illustrate them in Figure~\ref{fig:full}. We assume these functions would be standardized by the company, but may depend on specific data or project application and could be overridden by the user when necessary. 

\renewcommand\thefigure{\thesection.\arabic{figure}}    
\begin{figure*}[t!]
\includegraphics[width=.95\textwidth]{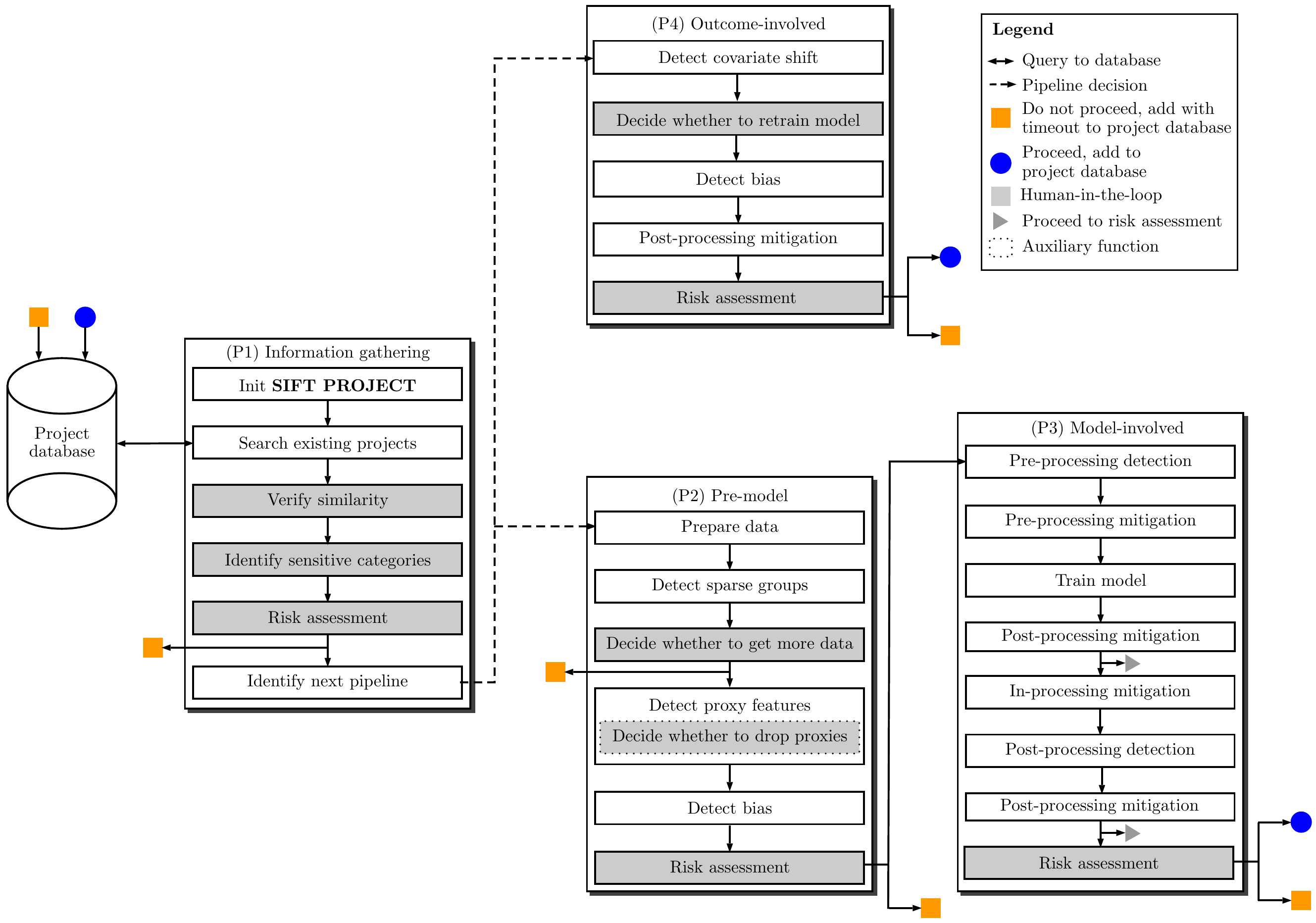}
\caption{A full diagram of the SIFT pipelines and functions.}
\label{fig:full}
\end{figure*}

\subsection{Information gathering}
\label{subsec:datacoll}

{\bf (1) Project initialization.}
The pipeline starts by initializing the {\bf\textsc{sift project}} object
which adds the prior model history (if any) obtained from the 
{\bf data location}, and any details on the data and
metadata. It initializes the {\bf\textsc{bias history}} object, which is
updated alongside bias-related steps taken throughout the project.

{\bf (2) Similar project identification.}  We now search the ML projects
database for similar projects, its location specifed by
{\bf db location}. The database can have a simple Web query interface
whence {\bf db location} would be a URL.  The search uses normal
information retrieval steps: remove non-alphanumeric characters,
normalize case, lemmatize words, remove stop words before vectorizing
the data and calculating a cosine similarity score. The SIFT user human-verifies the list of matched projects and adds relevant ones to
{\bf\textsc{sift project}.similar projects}. The user finds a relevant set of
sensitive features by considering those in {\bf\textsc{sift project}.similar projects}
and other factors (e.g., legal constraints and domain knowledge).  
The human step of sensitive category identification captures this action
and stores them as  {\bf\textsc{sift data}.sens features}. In absence of similar projects, external considerations can identify sensitive features.

{\bf (3) Preliminary risk assessment.}  The human-level risk assessment step now considers the information collected in the
current {\bf\textsc{sift project}} object, with the list of similar projects,
factoring in business, contractual, and legal constraints, as well as customer
impact. The {\bf\textsc{bias history}} object documents the outcome of the risk assessment.  If the decision is not to
proceed due to absence of fairness concerns or high risk of bias, the
project object is correspondingly updated. A timeout
component is set based on company guidelines and the project is added
to the project database before exiting SIFT. Else, the next pipeline
is determined---setting it to Outcome-involved if the model is deployed or to Pre-model
otherwise. A deployed model is one that is beyond model training and
development (for example if the model is field-trial ready or already in
production). Based on the pipeline determination, the
project is moved to either the Pre-model pipeline (Section 
\ref{subsec:premodel}) or the Outcome-involved pipeline (Section 
\ref{subsec:outcomeinvolved}).

\subsection{Pre-model pipeline}
\label{subsec:premodel}

{\bf (1) Data preparation.}  Standard practice in ML workflows, this step
involves data cleaning and feature engineering and is typically completed prior to performing any bias checks. While some aspects of
this could be mechanizable, data preparation often involves some input
and inspection of the data by the ML practitioner. The final data for the project should be updated and stored in the {\bf\textsc{sift data}} object.

{\bf (2) Sparse group detection.}
To train a fair ML model we need a sufficient number of training
samples for each of the subgroups defined by the sensitive attributes.
Otherwise, the ML model can have poor performance when predicting
results for samples of the under-represented subgroup in
practice. For example, Amazon abandoned an ML system intended to
automate the hiring process by identifying resumes of top technical
talent; its training on past resumes penalized female applicants due
to historical gender imbalance within the tech industry \cite{amazon}. If sparse groups are detected, the user can collect
additional data or terminate the project.  Else, the Model-involved pipeline will
attempt to address this issue using an algorithmic pre-processing
strategy such as reweighing or resampling~\cite{KamiranCalders2012}.

{\bf (3) Proxy feature detection.}  Removing sensitive attributes from the
set of features will not guarantee an unbiased ML model. One way
bias may remain in the data is through the existence of proxy
variables. For example, in the context of targeted advertising on
Facebook, \cite{Speicher2018potential} found that many features
provided on the Facebook ad platform were strongly correlated with
sensitive attributes like gender and race. SIFT checks for strong
pairwise correlations between sensitive and non-sensitive attributes. Pairwise correlation checks do not guarantee the removal of
all proxy variables: further bias checks are needed in the
Model-involved pipeline. In particular, when the number of sensitive
attributes and non-sensitive attributes is large, combinations of
non-sensitive attributes could create a proxy for a sensitive
attribute even when individual variables don't. Such multivariate
proxies would not be detected at this step. When univariate proxy
variables are detected, the SIFT user can drop the proxy variable from
consideration in later modeling steps.

{\bf (4) Marginalized group detection.}
Bias present in the target variable will be learned by the ML model. This step checks the target variable for marginalized groups to alert the SIFT user to this potential issue. If marginalized groups are detected, then an algorithmic mitigation
strategy can be implemented later in the Model-involved pipeline.

{\bf (5) Pre-model risk assessment.}  The last step of this pipeline asks
the SIFT user to perform a risk assessment given the
information learned in this pipeline---information that could fundamentally change
the project plan. For example, if a key input feature is found to
be a proxy for a sensitive attribute, the user may not wish to proceed. The {\bf\textsc{bias history}} object captures the steps
and results of each bias-related action taken in the pipeline. The
user reviews this, and {\bf\textsc{bias history}} documents the outcome of the risk assessment. If the user decides not to proceed, then project status is set to `Terminated', {\bf\textsc{sift project}.timeout} is set pursuant to company guidelines, and the project is added to the project database. Else, SIFT begins the Model-involved pipeline.  

\subsection{Model-involved pipeline}
\label{subsec:model}
Unlike other pipelines, the user may not proceed sequentially through all the steps in the standard flow given below. Constraints of time or computational resources may influence the choice and ordering of mitigation strategies. For example, if the ML model is computationally expensive to retrain and time-to-market is a concern, then the user may limit focus to only post-processing strategies. There is also no guarantee that any one mitigation strategy will resolve detected bias issues; multiple mitigation strategies may be required.  Further, a mitigation strategy may not exist that will address the source of bias, requiring designing of a novel mitigation strategy.

We thus allow for a choice of flow processes: standard or custom.  The enterprise would determine the appropriate {\it standard} flow that a majority of projects would follow.  For example, it could be set to closely follow the framework of \citet{dAlessandro2017}, which
works sequentially through the steps listed above.  The {\it custom}
flow allows the user control over the sequencing and implementation of
the steps of the pipeline.  The user can restrict attention to a
specific set of bias detection metrics and mitigation algorithms, copy a
routine from a similar project, or run a novel bias detection and
mitigation strategy designed for the application. The ability to copy
bias detection metrics and mitigation strategies used in similar ML
projects is a key feature of SIFT that helps reduce the cost of
reducing bias as more projects are added to the database.  The {\bf model flow} input to the {\bf\textsc{sift project}} object indicates the selected flow for the
project.

For reference, we provide examples of pre-, in-, and post-processing
mitigation algorithms, that have open-source code available
through~\cite{aif360}.

{\bf (1) Pre-processing mitigation.}  Pre-processing algorithms
\cite{Calmon2017,Feldman2015,KamiranCalders2012,Zemel2013} transform
the raw data to reduce if not remove bias. These algorithms address
bias in the raw data, for example, due to an under-representation of
samples from a protected group, and do not require access to the
training model or the model output. In case bias is detected, the system runs a pre-processing mitigation strategy. This function selects and implements the pre-processing strategy using the information collected in the Pre-model pipeline. It then returns the transformed dataset, and the pre-processing function, which is recorded in {\bf\textsc{bias history}}. The pre-processing function involves a transformation of the raw data, so the user will need to train the ML model regardless of prior model availability.

{\bf (2) Model training.}
This step of the pipeline train the ML model. All iterations in this model development process will be documented in {\bf\textsc{model history}} object.

{\bf (3) Model outcome bias detection.}
The existence of bias in the model outcome should be quantified using
one or more bias detection metrics~\cite{aif360}.  Often such metrics
check that the model outputs are equivalent across different values of a sensitive feature or demographic subpopulations. Examples of bias detection metrics include disparate impact, equalized odds, demographic parity, and statistical parity~\cite{aif360,Hardt2016,DaviesEtal17}.

{\bf (4) In-processing mitigation.}
In-processing algorithms incorporate one or more bias metrics directly into the prediction model and ensure that prediction accuracy is attained only under predefined fairness constraints stipulated by those bias metrics~\cite{Celis2018,Kamishima2012,Zhang2018}. After mitigation, SIFT returns information about the newly fitted model, which is stored as a new step in the {\bf\textsc{model history}} object, and the in-processing function, which is recorded in {\bf\textsc{bias history}}.  

{\bf (5) Post-processing mitigation.}
Post-processing algorithms \cite{KamiranKarim2012, Hardt2016, Pleiss2017} transform the outputs from a specific trained model and are model-agnostic in the sense that they do not require access to the training data or the trained model. These methods are particularly useful when there is a high cost for re-training the underlying model. SIFT selects and implements the post-processing mitigation strategy and subsequently returns the new predicted outcome and the selected post-processing mitigation function, which is recorded in {\bf\textsc{bias history}}.

{\bf (6) Model-involved risk assessment.}
The last step of this pipeline asks the SIFT user to perform a risk assessment given the results of this pipeline. Since no mitigation strategy is guaranteed to remove all forms of bias, this is a key step in the process and will depend on factors such as the extent to which the ML model might impact customers and the potential risk to the enterprise's brand image. This assessment should also take into account any degradation in utility due to bias mitigation steps. If the user decides to proceed, then SIFT will mark the project as scheduled for deployment and record the project in the project database. Otherwise, SIFT marks the project as terminated and records the project in the project database with the appropriate timeout specified.  In both cases, {\bf\textsc{bias history}} documents the risk assessment decision.

\subsection{Outcome-involved pipeline}
\label{subsec:outcomeinvolved}

As a typical example, this pipeline would involve: (1) Data change
detection, (2) Model outcome bias detection, (3) Post-processing
mitigation, and (4) Outcome-involved risk assessment. Steps 2 and 3 are
same as ones in Section~\ref{subsec:model}; details for steps 1 and 4
are provided below.

{\bf (1) Data change detection.}  Relevant variables may change over time,
triggering a check against recorded statistics about the model's
original training data for similarity~\cite{Mitchell19}.  For example,
features could be deleted, missing values or new categories could be
introduced, or there could be a distributional shift in one or more
variables. If an earlier model iteration that was trained on a
separate set of samples exists, then a typical example of this step
would check for a covariate shift in the feature variables. The prior data can be accessed through the list of pointers in {\bf\textsc{sift project}.older versions}.  If
necessary, summary statistics about the current data and prior data
can be compared if the full prior data set is not available.  It's important to note that checking for a covariate shift in the data may not detect all changes to the underlying data, and additional safeguards should be built-in that are specific to the company's applications. If a change to the underlying data is detected, then the user may decide to exit SIFT and retrain the model, or to continue through the pipeline to see if the change to the data results in a biased outcome.

{\bf (4) Outcome-involved risk assessment.}  Like previous pipelines, the
last step of the Outcome-involved pipeline asks the user to make a
final risk assessment, and {\bf \textsc{bias history}} object is updated to reflect the final decision. Any fairness
concerns that remain in the ML project must be weighed against the
cost and feasibility of developing, training, and deploying a new ML
model or working with an alternative third-party source. If the user
decides to proceed, then the project remains in deployment. Else,
the project is terminated and added to the project database with an
appropriate timeout specified.

\section{Simulation Details for Marketing Use Case}
\label{app:mrkt_data}

We use demographic data from the UCI Adult dataset and remove all
examples with missing information, resulting in $n = 45,222$ examples.
In our experiments, we use {\bf income}, {\bf sex}, and
\begin{itemize}[leftmargin=*, topsep=1pt]
\setlength\itemsep{0em}
\item {\bf age} - binned as $[17, 25]$, $[26, 35]$, $[36, 45]$, $[46, 55]$, $[56, 65]$, $[66, 75]$, or $75+$, and converted to its one hot encoding,
\item {\bf marital status} - converted to ``married'' or ``single'',
\item {\bf race} - converted to ``white'' or ``non-white''.
\end{itemize}
We treat \{{\bf marital status}, {\bf race}, {\bf sex}\} as the set of sensitive features.

To generate consumer segments that are correlated with the sensitive
features, we simulate $C^c_{i,j} \sim Bin(\tilde{p_i})$ for $j = 1,
\ldots, 5$ and $i = 1, \ldots, n$, where
\[
\tilde{p_i} = \frac{\exp\left(-1 + I_m + I_s\right)}
  {1 + \exp\left(-1 + I_m + I_s \right)},
\]
with
\begin{align*}
    I_m &= \mathbbm{1}_{\{marital\_status_i = Married\}},\\
    I_s &= \mathbbm{1}_{\{sex_i = Male\}}.
\end{align*}
In addition, we simulate consumer segments $C^u_{i, j} \sim Bin(p_j)$
for $j = 1, \ldots, 45$ and $i = 1, \ldots, n$, where $p_j \sim U(0.2,
0.8)$, to be consumer segments that are uncorrelated with the
sensitive features.

We define
\(
\mathbf{X} = [\text{age}, \text{income}, \mathbf{C^c}, \mathbf{C^u}]
\)
to be the set of features for model training.  To simulate the target
variable, $\mathbf{y}$, we define $\boldsymbol \beta$ to be a vector
of coefficients corresponding to the features in $\mathbf{X}$.  We
simulate coefficients for income, $\mathbf{C^c}$, and the first 10
features in $\mathbf{C^u}$ from $U(-2, 2.5)$.  We set all other
coefficients to zero.  Then
$y_i \sim Bin(p^y_i)$ for $i = 1, \ldots, n$, where
\[
p^y_i = \frac{\exp(-0.5 + \mathbf{X}_i \boldsymbol \beta + z_i)}
{1 + \exp(-0.5 + \mathbf{X}_i \boldsymbol \beta + z_i)}.
\]
Here $z_i \sim N(0, 1)$, for $i = 1, \ldots, n$, prevents a perfect model fit.